\documentclass[a4paper,fleqn]{cas-dc}
\usepackage[authoryear,longnamesfirst]{natbib}
\usepackage{amsmath,amssymb,amsfonts}
\usepackage{algorithm,algorithmic}
\usepackage{graphicx}
\usepackage{booktabs}

\def\tsc#1{\csdef{#1}{\textsc{\lowercase{#1}}\xspace}}
\tsc{WGM}
\tsc{QE}

\begin{document}

\let\WriteBookmarks\relax
\def\floatpagepagefraction{1}
\def\textpagefraction{.001}

\title{Latent Alignment with Deep Set EEG Decoders}

\shorttitle{Latent Alignment with Deep Set EEG Decoders}
\shortauthors{Bakas et al.}

\author[1, 2, 4]{Stylianos~Bakas}[orcid=0000-0003-1054-0169]
\fnmark[1]
\cormark[1]
\ead{simpakas@csd.auth.gr}

\author[1, 4]{Siegfried~Ludwig}[orcid=0000-0001-9693-9482]
\fnmark[1]

\author[1, 4]{Dimitrios~A.~Adamos}[orcid=0000-0001-6700-1057]

\author[1, 2, 4]{Nikolaos~Laskaris}[orcid=0000-0002-1960-394X]

\author[1, 3, 4]{Yannis~Panagakis}[orcid=0000-0003-0153-5210]

\author[1, 4]{Stefanos~Zafeiriou}[orcid=0000-0002-5222-1740]

\fntext[1]{equal contribution}

\affiliation[1]{
    organization={Department of Computing},
    addressline={Imperial College London},
    postcode={SW7 2RH},
    country={United Kingdom}}
\affiliation[2]{
    organization={School of Informatics},
    addressline={Aristotle University of Thessaloniki},
    postcode={54124},
    country={Greece}}
\affiliation[3]{
    organization={Department of Informatics and Telecommunications},
    addressline={National and Kapodistrian University of Athens},
    postcode={15784},
    country={Greece}}
\affiliation[4]{
    organization={Cogitat Ltd.},
    city={London},
    country={United Kingdom}}

\begin{abstract}
    The variability in EEG signals between different individuals poses a significant challenge when implementing brain-computer interfaces (BCI). Commonly proposed solutions to this problem include deep learning models, due to their increased capacity and generalization, as well as explicit domain adaptation techniques. Here, we introduce the Latent Alignment method that won the Benchmarks for EEG Transfer Learning (BEETL) competition and present its formulation as a deep set applied on the set of trials from a given subject. Its performance is compared to recent statistical domain adaptation techniques under various conditions. The experimental paradigms include motor imagery (MI), oddball event-related potentials (ERP) and sleep stage classification, where different well-established deep learning models are applied on each task. Our experimental results show that performing statistical distribution alignment at later stages in a deep learning model is beneficial to the classification accuracy, yielding the highest performance for our proposed method. We further investigate practical considerations that arise in the context of using deep learning and statistical alignment for EEG decoding. In this regard, we study class-discriminative artifacts that can spuriously improve results for deep learning models, as well as the impact of class-imbalance on alignment. We delineate a trade-off relationship between increased classification accuracy when alignment is performed at later modeling stages, and susceptibility to class-imbalance in the set of trials that the statistics are computed on.
\end{abstract}



\maketitle

\section{Introduction}
\label{sec:introduction}

Subject independent EEG decoding models have long been a sought-after goal for enabling the translation of brain-computer interfaces from research into practice. Previous systems relied on extensive calibration sessions for each new subject, which is a tedious process and presents a significant hurdle for more wide-spread use \cite{rashid2020current, padfield2019eeg}.

The need for novel approaches in EEG subject adaptation spurred the BEETL competition at the Conference on Neural Information Processing Systems (NeurIPS) in 2021, which aimed to identify novel approaches for addressing subject-independence and training across heterogeneous datasets \cite{wei20222021}. This paper introduces the Latent Alignment method we developed for the winning competition entry.

Performing feature standardization represents a common class of methods for unifying covariate shifts between subjects that rely on statistical distribution estimates computed on a context set of trials. These are most often applied on the input signal \cite{apicella2023effects, congedo2017riemannian, he2019transfer}, but the personalized standardization of classification features has also been performed \cite{fdez2021cross, jimenez2020custom}. We expect that aligning within the latent space will improve accuracy, because it will be more relevant to the classification task.

Understanding Latent Alignment as a permutation equivariant function applied on the context set of trials from a given subject allows us to reformulate the resulting deep learning model as a deep set architecture \cite{zaheer2017deep}. To the best of our knowledge this is the first time deep sets are encountered in the literature on EEG decoders. Utilizing the additional information contained in the context set of trials, rather than processing each trial independently, holds promise for improved EEG decoding performance. We compare the classification accuracy of the proposed method to other statistical alignment techniques.

Considering the practical application of such approaches, we assume that an unbalanced class distribution would reduce their effectiveness, as it will impact the statistical estimates computed on the trial context set. This effect is expected to be more pronounced if alignment is performed in later stages of the model, where different classes are more linearly separable.

Aiming to mitigate this problem of class imbalance in the context set, we hypothesize that training a model on randomized class distributions, while applying statistical alignment on the latent model stages, reduces the susceptibility to class imbalance encountered during inference.

All top three teams in the BEETL competition relied on deep learning approaches, which promise better transferability across subjects due to their increased capacity. A recent review by Craik et al. found that a majority of studies using deep learning on EEG did not mention or chose not to perform explicit artifact removal \cite{craik2019deep}. This is likely due to the assumption that such models are more robust to noisy signals. We caution however that omitting the artifact removal step can lead deep learning models to associate class-discriminative artifacts, such as saccadic eye movements following the cue presentation, with the classification label \cite{frolich2015investigating}. Appropriate steps need to be taken to ensure uncontaminated results.

In summary, the contributions of this paper are as follows:
\begin{itemize}
    \item We introduce the Latent Alignment approach for subject-independent models and its formulation as a deep set.
    \item We compare the proposed Latent Alignment method to recent statistical adaptation techniques on a multitude of classification tasks and deep learning models.
    \item We study the impact of class imbalance on performance and its relation to the latent space that alignment is executed within.
    \item We show the impact of class-discriminative artifacts when training deep learning models on EEG data.
\end{itemize}

\section{Related Work}
\label{sec:related_work}

Aligning distributions within the classification space resulting from the feature extraction steps, rather than simply standardizing the temporal EEG signal, holds promise for improved subject transferability.

Among the state-of-the-art methods for subject adaptation are Riemannian techniques, which perform standardization of subject-wise distributions on the semi-positive definite (SPD) manifold of EEG covariance matrices \cite{congedo2017riemannian, zanini2017transfer, yger2016riemannian, lotte2015signal}. Performing classification on the Riemannian manifold requires specialized models however, and cannot straightforwardly be adapted for deep learning techniques, although there are recent efforts such as SPDNet \cite{huang2017riemannian}, also applied to EEG decoding \cite{hajinoroozi2017driver}. We focus here on statistical alignment techniques that are broadly compatible with existing deep learning models.

Euclidean alignment aims to adapt Riemannian alignment techniques for standard deep learning models, such as convolutional neural networks \cite{he2019transfer}. This approach performs spatial whitening using the Euclidean mean of spatial covariance matrices across multiple trials of a given subject. The aligned signals are then provided as input to the deep learning model.

Alignment on the SPD manifold can be seen as aligning within the classification space for some conventional methods, since these would directly be applied on the resulting manifold, which includes the signal powers. Deep learning methods however perform additional steps of feature extraction as part of the model architecture, which results in a discrepancy between the alignment space and the final classification space.

Adaptive Batch Normalization (Adaptive BatchNorm) is a deep learning domain transfer technique developed in the image domain \cite{li2018adaptive}, which was introduced to EEG decoding by \cite{jimenez2020custom, xu2021improving}. This approach simply replaces the statistics from the source dataset applied in every batch normalization layer with statistics obtained from the target dataset. Compared to Euclidean alignment it has the advantage of applying its adaptation step in the latent classification space of a deep learning model instead of the input. During the training stage however, this approach just performs standard batch normalization without considering the inter-subject variability within the training set.

Deep sets are a deep learning framework developed for the classification of sets, rather than fixed dimensional vectors \cite{zaheer2017deep}. For a function to operate on a set it must be permutation invariant (or equivariant) with respect to an undefined number of input samples. An invariant deep set produces a single output from a set of input samples, while an equivariant deep set results in the same number of outputs as there are input samples, each transformed by the invariant set function. The authors propose to use sum or maxpooling as a commutative function to aggregate the information in the set.

\section{Methods}
\label{sec:methods}

\subsection{Alignment Techniques}

In order to employ the alignment techniques studied here, we carefully compose each batch of training data to include a fixed number of trials from each included subject session. Depending on the analysis, these batches include either balanced or randomly unbalanced class distributions of trials for each subject. Alignment is then straightforwardly performed using the subject-wise statistics obtainable from each batch. During inference, all unlabeled trials of individual subjects are decoded simultaneously, using the full statistics. The class distributions then reflect the distribution in the respective dataset.

All three alignment methods obtain the relevant statistics by averaging across trials as well as across the time dimension of each trial, as is custom with batch normalization. In the following notation, we omit the time dimension for the sake of brevity.

\subsubsection{Latent Alignment}

Intuitively, Latent Alignment can be understood as a subject-wise batch normalization, applied during training and inference. A preliminary version of the technique was used in our winning entry to the BEETL NeurIPS competition \cite{wei20222021, bakas2022team}. We will first introduce its formulation as distribution standardization, followed by the deep set formulation.

The alignment context set is constructed as a batch of $n$ trials from the current subject. A model forward pass is then performed on the batch. At each alignment layer, the mean $\Bar{x}_{sbj} \in \mathbb{R}^{d}$ and standard deviation $s_{sbj} \in \mathbb{R}^{d}$ across trials with latent features $x_i \in \mathbb{R}^{d}$, $i \in \{1, ..., n\}$ of dimensionality $d$ are computed as
\begin{equation}
\begin{split}
    \Bar{x}_{sbj} &= \frac{1}{n} \Sigma_i^n x_i \\
    s_{sbj} &= \sqrt{\frac{\Sigma_i^n (x_i - \Bar{x}_{sbj})^2}{n - 1}},
\end{split}
\end{equation}
treating each feature dimension independently. The resulting statistics are then applied to standardize the latent distribution across subject trials in the batch $\textbf{x} \in \mathbb{R}^{n \times d}$, such that
\begin{equation}
    \textbf{y} = \frac{\textbf{x}-\Bar{x}_{sbj}}{s_{sbj}} \cdot \alpha + \zeta,
\end{equation}
resulting in a batch of trials with standardized latent features $\textbf{y} \in \mathbb{R}^{n \times d}$. Trainable scale $\alpha \in \mathbb{R}^{d}$ and bias $\zeta \in \mathbb{R}^{d}$ parameters, shared across all subjects, are applied on the feature distributions, as is commonly performed in batch normalization \cite{ioffe2015batch}. The aligned features are then propagated to the next trainable weight layer of the deep learning model, after which the alignment procedure is repeated.

The Latent Alignment approach can alternatively be expressed as a deep set \cite{zaheer2017deep}. The function $\rho:\mathbb{R}^{n \times d}\to\mathbb{R}^{n \times d}$, which is permutation equivariant with respect to the set of $n$ input trials, is applied on the  latent features coming from the previous layer. The resulting standardized features are transformed by a trainable weight matrix $\Gamma \in \mathbb{R}^{d \times d'}$, mapping from $d$ input to $d'$ output features, and bias $\beta \in \mathbb{R}^{d'}$, such that
\delimitershortfall=-0.5pt  
\begin{equation}
\begin{split}
    \rho(\textbf{x}) &= \textbf{x} + \frac{\textbf{x} - \Bar{x}_{sbj}}{s_{sbj}} \cdot \alpha + \zeta\\
    f(\textbf{x}) &= \sigma(\beta + (\textbf{x} - \rho(\textbf{x}))\Gamma),
\end{split}
\end{equation}
where $\sigma:\mathbb{R}^{d'}\to\mathbb{R}^{d'}$ is a non-linear activation function. It is interesting to note that in this view, any deep learning model that applies batch normalization can be seen as a deep set. In that case however, the set readout will be approximately constant across all trials and subjects, representing the latent statistics of the training dataset, which makes it a trivial case.

When applying Latent Alignment during inference, we include each test trial in the context set, which is possible because the Latent Alignment method does not require class labels. When performing inference on a single new trial, the previously obtained context set can be concatenated to form a growing batch of trials, on which the new statistics can be computed.

Latent Alignment repeatedly standardizes feature distributions following successive layers of feature extraction in a deep learning model, up to and including the final classification space. Following our hypothesis that performing alignment within the classification space rather than only the input, this approach should be optimal. Furthermore, applying alignment both during training and inference proactively addresses inter-subject variability in the training set, and eliminates the difference between training and inference behaviour.

\subsubsection{Euclidean Alignment}

Euclidean Alignment \cite{he2019transfer} standardizes the distribution of EEG input signals $\textbf{x} \in \mathbb{R}^{n \times d}$ by performing spatial whitening with the average spatial covariance matrix $\Sigma_{sbj} \in \mathbb{R}^{d \times d}$. Alignment of the trials is performed with
\begin{equation}
    \textbf{y} = \Sigma^{-\frac{1}{2}}_{sbj}\textbf{x},
\end{equation}
where we compute the square root of the covariance matrix using the Cholesky decomposition, followed by the matrix inverse. The aligned signals $\textbf{y} \in \mathbb{R}^{n \times d}$ are then passed into the deep learning model. For numerical stability we recenter signals per electrode and rescale them using the total trial standard deviation across electrodes (average global field power) before computing the covariance matrix.

\subsubsection{Adaptive BatchNorm}

The Adaptive BatchNorm approach applies simple batch normalization layers in the training stage and then replaces the statistics of every normalization layer with statistics obtained on a set of trials from the target subject. We implement this by running simple batch normalization layers during training and then utilizing the subject-specific trials in each composed batch to obtain personalized statistics during inference.

\subsection{Models and Datasets}
\label{sec:datasets}

Three datasets representative of the variety of classification tasks met in EEG BCI applications were selected to test our Latent Alignment approach. On each dataset, a well-established deep learning model developed specifically for the respective task is trained with varying alignment methods. We added a batch normalization layer on the input for each model, which applies batch normalization per electrode without trainable weight and bias, which is then used for Adaptive BatchNorm or Latent Alignment depending on the experiment.

All models are trained using the Adam optimizer \cite{kingma2014adam} with learning rate 1e-3, weight decay 1e-3, dropout 0.25 \cite{srivastava2014dropout}. On all tasks we perform subject-independent cross-validation with 10 folds, such that the subjects in the respective validation sets were not seen during training of the model.

\begin{figure}[tbp]
    \centering
    \includegraphics[width=\columnwidth]{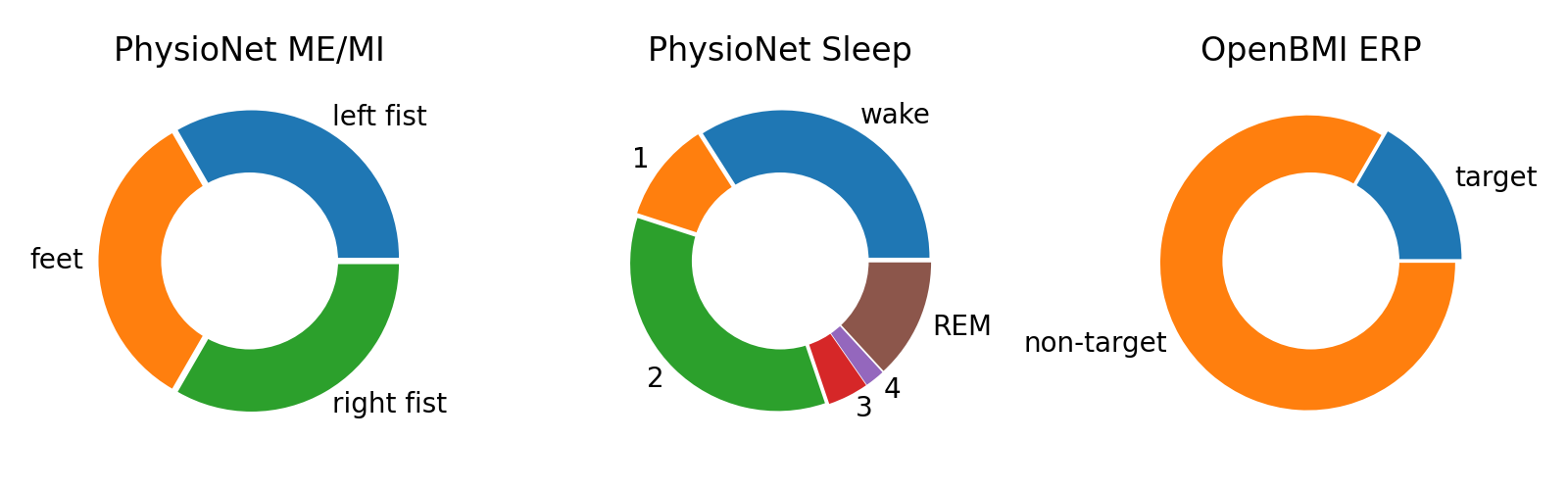}
    \caption{Class distributions for the datasets used in this study.}
    \label{fig:distributions_piechart}
\end{figure}

\subsubsection{PhysioNet MI}

The first dataset is PhysioNet MI, which contains recordings of 109 subjects performing motor execution (ME) and imagery (MI) of hands and feet \cite{goldberger2000physiobank, schalk2004bci2000}. We drop six subjects due to the inconsistent number of trials and sampling rates \footnote{subjects 88, 89, 92, 100, 104, 106}. The dataset contains EEG recordings from 64 electrodes at a sampling rate of 160Hz, cut into trials of length 4.1 seconds. We apply a third-order zero-phase Butterworth bandpass filter in the range 4-40Hz and a Notch filter at 60Hz, as well as common average re-referencing.

Three classes are chosen, including left fist, right fist and both feet. We employ EEGNet, one of the most established deep learning models to date on the motor decoding task \cite{lawhern2018eegnet}. It is parameterized by 8 temporal and 2 spatial filters for a total of 16 channels, and we adapt the convolutional kernel sizes to match the sampling rate of 160Hz. On each task, the model is trained for 100 epochs with each training batch consisting of 4 subjects with 12 trials per subject for a total batch size of 48.

\subsubsection{PhysioNet Sleep}

The PhysioNet Sleep Cassette dataset includes 78 subjects with EEG recordings from 2 bipolar electrodes (Fpz-Cz and Pz-Oz) \cite{goldberger2000physiobank, kemp2000analysis}. We discard any data from the awake condition except for the 30 minutes before and after sleep. Due to the very small number of trials in deep sleep stages 3 and 4 (Figure \ref{fig:distributions_piechart}), we combine these classes into one. The dataset is cut into trials of length 30 seconds. The sampling rate of the dataset is 100Hz, and we apply a third-order zero-phase Butterworth bandpass filter in the range 0.1-45Hz.

For the sleep stage classification task we use the well established DeepSleep model \cite{chambon2018deep}. Batch normalization layers are added after each of the three convolutional layers. We collapse the latent channels of 8 spatial and 2 temporal filters into the same dimension for a total of 16 channels to allow for computation to occur across these channels. The model is trained for 5 epochs with 64 trials per subject session in each batch for a total batch size of 256. To counteract the imbalance of classes, the loss of each trial is weighted with the inverse of the class distribution in the training dataset.

\subsubsection{OpenBMI ERP}

The last task involves the classification of P300 event-related potential (ERP) responses on the OpenBMI dataset \cite{lee2019eeg}. It includes EEG recordings with 62 electrodes from 54 subjects, each including 2 sessions with 2 phases. We drop non-standard electrodes\footnote{electrodes TP9, TP10, PO9, PO10, FT9, FTT9h, TTP7h, TPP9h, FT10, FTT10h, TPP8h, TPP10h, F9, F10}, leaving 48 electrodes. The dataset is cut into trials of 1 second as proposed by the authors of the dataset, resampled to 100Hz and a third-order zero-phase Butterworth filter is applied in the range 0.5-45Hz. This is followed by common average re-referencing.

In this task the EEGInception model is employed with 8 temporal and 2 spatial filters, which taken together with the 3 parallel inception depths amounts to 48 channels \cite{santamaria2020eeg}. The model is trained for 10 epochs with 12 trials per subject session in each batch for a total batch size of 48. All experiments on this dataset are performed with batches containing the fixed class ratio of 5 non-target trials for every target trial, representing the dataset average. To counteract the imbalance of classes, the loss of each trial is weighted with the inverse of the class distribution in the training dataset.

\section{Results}
\label{sec:results}

We first examine trained models for contamination with class-discriminative artifacts by investigating their spatial filters. This is followed by the comparison of the different alignment methods and finally the analysis of the impact of class distributions on subject alignment, including comparisons between training on class-balanced and unbalanced data.

\subsection{Spatial Filter Analysis}

\begin{figure*}[tbp]
    \centering
    \includegraphics[width=\textwidth]{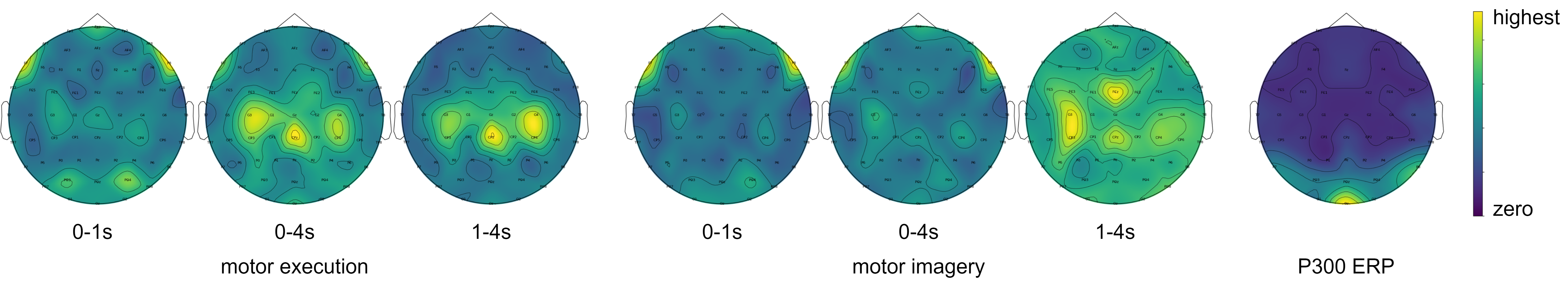}
    \caption{Topographic representation of the respective trained spatial filter layer for the motor and ERP classification tasks, with models trained on different parts of the trial for the motor task. Models trained on the first second exhibit a clear focus on F7, F8 and Fpz electrodes, while models trained on the rest of the trial exhibit a focus on motor-related central electrodes. Each representation shows mean absolute values of the spatial filter weights averaged across filter channels and cross-validation folds to display the grand average electrode relevance. Three electrodes were omitted for visual clarity (T9, T10, Iz).}
    \label{fig:spatial_topomaps}
\end{figure*}

Since deep learning models trained on EEG signals without artifact removal could utilize class-discriminative artifacts for classification, we investigated the linear spatial weights of baseline EEGNet models trained on different parts of the trial (Figure \ref{fig:spatial_topomaps}). It can be observed that models trained on the first second place a focus on F7, F8 and Fpz electrodes. This suggests that the decoder has learned to exploit eye movement artifacts, which inherently carry class-relevant information due to the adopted cue-based experimental design. Models trained on the remaining three seconds of the trial focus on C3, CP3, CPz, C4, CP4 for the motor execution task, and C3, CP3, FCz, CPz and CP4 for motor imagery, which are in the expected locations over the motor cortex.

\begin{figure}[tbp]
    \centering
    \includegraphics[width=\columnwidth]{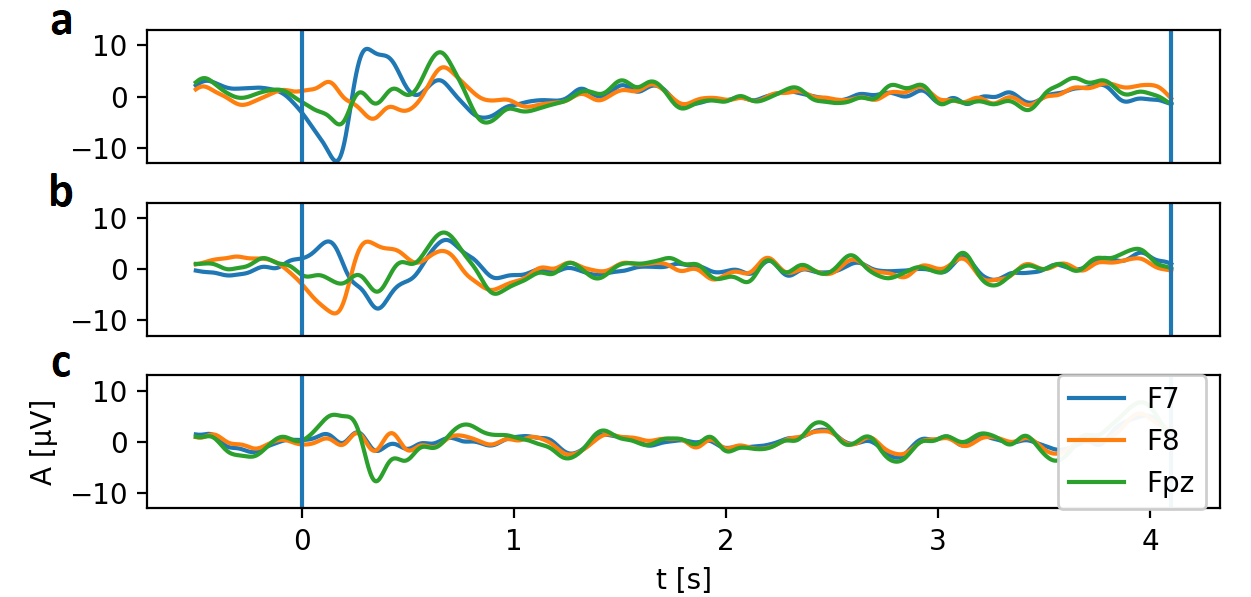}
    \caption{Grand average EEG amplitude across subjects for select electrodes containing eye activity on Physionet executed movement trials. Bandpass filtered in 1-8Hz. Clear differences between classes are apparent during the first second of the trial, after the cue appears on the screen. a) Left hand, b) right hand, c) feet.}
    \label{fig:physionet_eeg}
\end{figure}

Plotting averaged EEG signals from PhysioNet motor trials indeed reveals systematic eye movements, presumably following the task presentation on the screen (Figure \ref{fig:physionet_eeg}). The polarity of the eye movement artifact in F7 and F8 is reversed between left and right hand trials, supporting this interpretation. While details on the dataset are not available, we assume that an instruction stimulus for moving the left hand was presented on the left side of the screen at trial time zero, which the subject tended to follow with a saccadic eye movement. For the right hand and feet classes the cue would have been on the right side or lower part of the screen, respectively. The deep learning model can then classify the task based on the systematic eye movement artifacts in the EEG signal. The amplitude of these artifacts is expected to be proportional to the angle of the saccadic eye movement following the visual stimulus.

\begin{table}[tbp]
\centering
\caption{Classification results for motor tasks when training the EEGNet baseline model on different parts of the trial window. Models trained on the first second have much higher classification accuracy than models trained on the rest of the trial. Reported as balanced accuracy, mean (std).}
\label{tab:physionet_windows}
\begin{tabular}{@{}llll@{}}
\toprule
    & 0-1s          & 0-4s          & 1-4s          \\ \midrule
ME  & 0.597 (0.056) & 0.606 (0.060) & 0.521 (0.060) \\
MI  & 0.582 (0.048) & 0.567 (0.057) & 0.450 (0.035) \\ \bottomrule
\end{tabular}
\end{table}

The strong contribution of these artifacts towards the classification accuracy can be seen in Table \ref{tab:physionet_windows}, especially on the motor imagery paradigm. When classifying the signal using only the first second of the trial, which contains the eye movement artifact, the model gets 58.2\% accuracy, while it only reaches 45.0\% accuracy when classifying the other three seconds excluding the first.

For any further analysis on the motor tasks we exclude the first second of the trial during model training and testing to avoid the contribution of systematic eye movements to the classification.

Spatial filters of baseline models trained on the P300 speller ERP classification task exhibit a clear focus on Oz (Figure \ref{fig:spatial_topomaps}), with some contribution from neighbouring electrodes up to P7, P8 and Pz. This topography is surprising, since visual speller paradigms target the P300 response, which is located in the central-parietal region \cite{rezeika2018brain}. In the OpenBMI dataset used here, the authors introducing the dataset aimed to enhance the ERP response by drawing a human face stimulus over the letter when flashing it, which would contribute to differences in brain activity associated with visual processing between target and non-target stimuli. This is also evidenced in the topographical data validation the authors of the dataset performed, which in addition to the expected P300 around Cz shows strong differences in the occipital region \cite{lee2019eeg}.

\subsection{Classification Performance}

Classification results are given in Table \ref{tab:results}. Given that our cross-validation folds were created with controlled random seed and therefore identical across experiments, we performed paired t-tests when comparing classification accuracy. All alignment methods had the highest impact on the motor classification tasks, and the least impact on sleep stage classification. Paired t-tests across validation folds between the respective baseline models and the alignment techniques reveal that while all alignment methods lead to statistically significant increases on the motor tasks, only Latent Alignment lead to a statistically significant increase in performance on the sleep stage task ($p < 0.05$). Regarding P300 ERP classification, both Adaptive BatchNorm ($p < 0.05$) and Latent Alignment ($p < 0.01$) showed significant increases in balanced accuracy.

\begin{table*}[tbp]
\centering
\caption{Comparison of the baseline and alignment techniques on different EEG classification tasks with their respective deep learning models. The motor tasks are trained either on class-balanced or -unbalanced batches, but always tested on balanced distributions. Results are reported as balanced accuracy, mean (std) across folds. Paired t-test of alignment methods compared to the baseline:\\
$\dagger$ $p < 0.10$, * $p < 0.05$, ** $p < 0.01$, *** $p < 0.001$.}
\label{tab:results}
\begin{tabular}{@{}lllll@{}}
\toprule
                & Baseline       & Euclidean Alignment  & Adaptive BatchNorm  & Latent Alignment           \\ \midrule
EEGNet: ME (balanced)   & 0.521 (0.060)  & 0.625 (0.057)***     & 0.630 (0.055)***    & \textbf{0.641} (0.043)***  \\
EEGNet: ME (unbalanced) & 0.531 (0.054)  & 0.627 (0.047)***     & 0.624 (0.041)***    & \textbf{0.635} (0.057)***  \\ \midrule
EEGNet: MI (balanced)   & 0.450 (0.035)  & 0.483 (0.050)*       & 0.508 (0.039)***    & \textbf{0.517} (0.054)***  \\
EEGNet: MI (unbalanced) & 0.438 (0.036)  & 0.489 (0.046)**      & 0.512 (0.046)***    & \textbf{0.523} (0.052)***  \\ \midrule
DeepSleep: Sleep Stages    & 0.732 (0.031)  & 0.725 (0.020)        & 0.731 (0.031)       & \textbf{0.749} (0.024)*    \\ \midrule
EEGInception: P300 ERPs       & 0.854 (0.026)  & 0.861 (0.021)        & 0.866 (0.020)*      & \textbf{0.869} (0.022)**   \\ \bottomrule
\end{tabular}
\end{table*}

For the classification of motor execution and imagery trials we distinguish between models trained on class-balanced and randomly unbalanced batches. We observe that all alignment methods lead to a statistically significant increase in the classification of motor execution trials ($p < 0.001$) of about 9-12\%. For motor imagery, all methods lead to an increase of about 3-8\%, albeit with reduced significance levels for Euclidean Alignment ($p < 0.05$ with class-balanced training and $p < 0.01$ with class-unbalanced training for Euclidean Alignment; $p < 0.001$ for Adaptive BatchNorm and Latent Alignment).


We expected that aligning within the latent classification space would lead to increased classification accuracy. The results confirm that the proposed Latent Alignment technique consistently produces the highest performance on all tested tasks, albeit at times with a small margin of difference to the other methods. Consistently second is Adaptive BatchNorm, followed by Euclidean Alignment.

\subsection{Latent Distributions}

\begin{figure}[!t]
    \centering
    \includegraphics[width=\columnwidth]{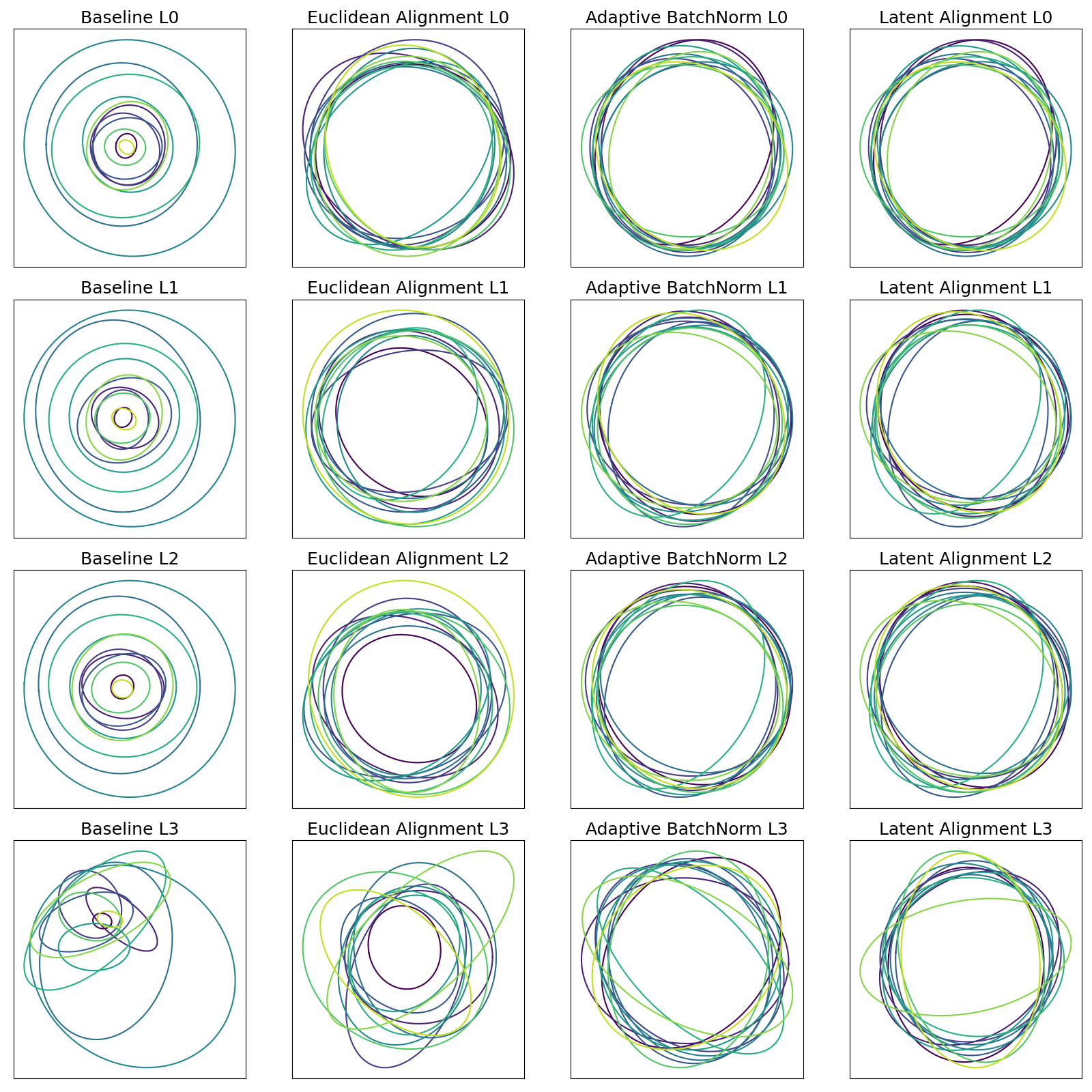}
    \caption{Latent distributions of different subjects at various stages in the EEGNet model trained on the motor execution task, comparing different alignment methods. Shown are the first two components after MDS dimensionality reduction following each of the four batch normalization or alignment layers. An ellipse is drawn around the 95\% confidence interval of the latent distribution of all trials for each of eleven subjects taken from the first fold validation set. The classes are therefore balanced within each distribution.}
    \label{fig:latent_space}
\end{figure}

A visualization of latent data distributions at the various stages of processing throughout the EEGNet model trained on the motor execution task can be seen in Figure \ref{fig:latent_space}. When compared to the baseline model, which exhibits strong inter-subject variability, we observe that every alignment method tested in this study noticeably reduces the inter-subject variability, both on the model input (L0) as well as in the latent stages (L1-L3). Euclidean Alignment is only applied on the input signal, which explains the increasing divergence between subject distributions toward the later stages of the model. In contrast, both Adaptive BatchNorm and Latent Alignment are applied repeatedly throughout the model, and subject distributions therefore remain well aligned. This observation agrees with the increased classification accuracy of Adaptive BatchNorm and Latent Alignment when compared to Euclidean Alignment.

\subsection{Impact of Class Distributions}

\begin{figure*}[tbp]
    \centering
    \includegraphics[width=\textwidth]{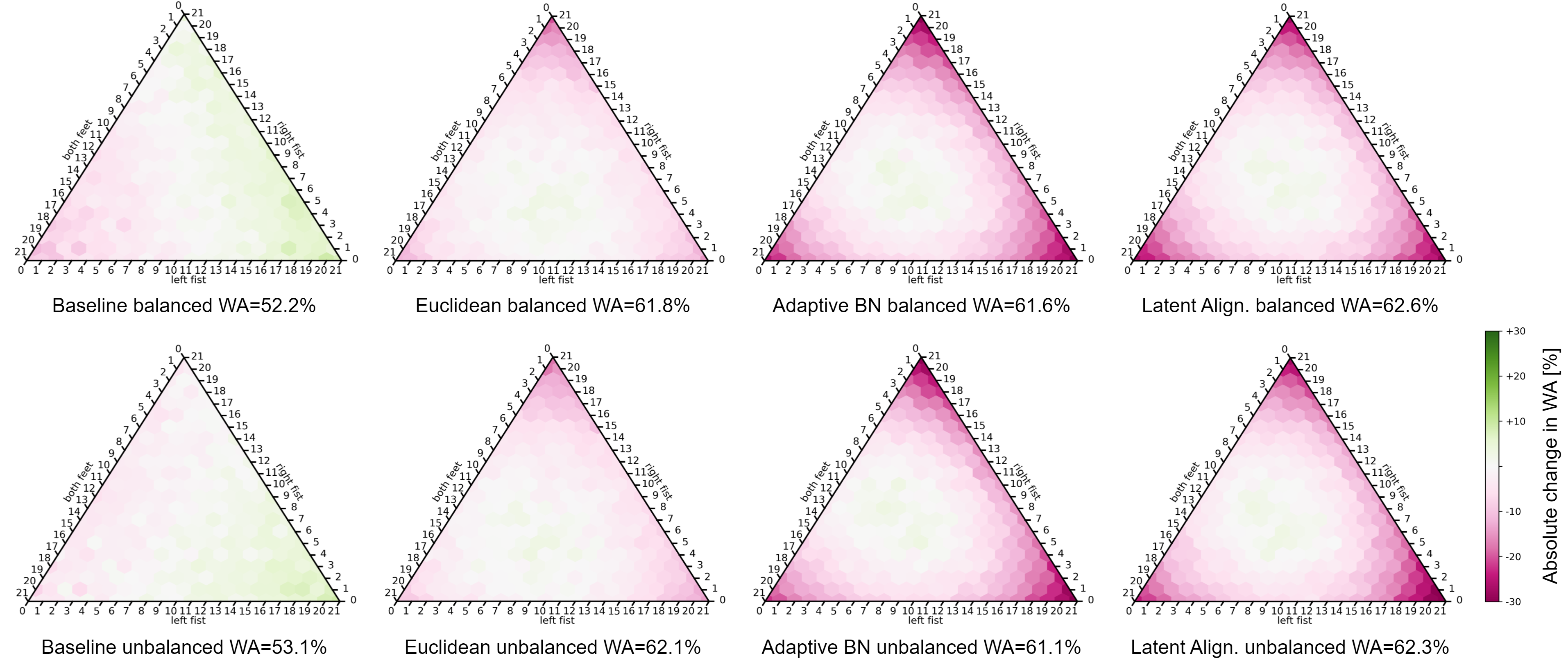}
    \caption{Evaluation of the impact of class distributions on classification accuracy for the different alignment methods, averaged over fold validation sets on the motor execution task. The top row shows results when trained on class-balanced batches, while the bottom row shows results trained on randomly unbalanced batches. Aggregate results are given in weighted accuracy (WA), using the multinomial probability distribution of random class imbalances for a context size of 21 trials to weight accuracy numbers.}
    \label{fig:triangle_plots}
\end{figure*}

To analyze the impact of varying class distributions on statistical alignment methods, we obtained the classification accuracy for every possible combination of the three classes in a context size of 21 trials (Figure \ref{fig:triangle_plots}). As an aggregate performance metric over this set of accuracy numbers, we weighted each outcome by its respective probability given a multinomial distribution over possible random compositions of the context, and term this "weighted accuracy (WA)" for the purpose of this study. Specifically, the WA for three classes, with the number of trials in each class given as $(i, j, k)$, is calculated as
\begin{equation}
    \text{WA} = \sum_i^n \sum_j^{n-i}
    \frac{n!}{i!j!k!} p_i^i p_j^j p_k^k
    \cdot acc_{ijk},
\end{equation}
where $k=n-i-j$ and in our case $p_i=p_j=p_k=\frac{1}{3}$ and $n=21$.

We assumed that class imbalance in the context set of trials would reduce the effectiveness of statistical domain adaptation techniques. The distribution analysis highlights the significant fall-off in classification accuracy for severely imbalanced distributions, reaching close to random level performance in cases where all trials in the context set belong to the same class. The relative robustness of Euclidean Alignment to class imbalance is likely due to it being applied only on the input signal, which is less linearly discriminative between classes. We therefore conjecture that the more linearly separable the classes are in a distribution, the more susceptible to class imbalances the statistical alignment methods are. This means that while applying alignment at later stages in the model leads to higher accuracy, it also leads to stronger requirements for approximate class balance in the trial context set.

Our hypothesis states that applying Latent Alignment during training and using class-unbalanced batches would result in models more robust to skewed class distributions on the alignment context set. However, we found that training on batches with random class balance did not result in significantly different weighted accuracy relative to balanced batches (paired t-test over 10 folds; Baseline: $t$=1.085, $p$=0.306; Euclidean Alignment: $t$=0.363, $p$=0.725; Adaptive BatchNorm: $t$=-0.450, $p$=0.664; Latent Alignment: $t$=-0.843, $p$=0.421).

\section{Discussion}
\label{sec:discussion}

We identified contamination with class-discriminative eye movements in the first second of trials in the motor task and therefore dropped this segment from further analysis. This step might however drop relevant motor-related brain activity. Instead, artifact removal techniques could be applied in order to retain the contaminated window of each trial as part of the model input. The most commonly employed approaches for EEG artifact removal include independent component analysis and linear regression \cite{uriguen2015eeg, jiang2019removal}. Validation steps, such as performed in this study by visualizing the trained spatial filters of decoding models, should nevertheless be performed to ensure uncontaminated results. More interpretable decoding models would benefit with this type of analysis \cite{ludwig2021eegminer}.

The Euclidean Alignment technique performed reasonably well when compared to other techniques that are more native to deep learning models. Its reliance on spatial covariance matrices, however, likely contributed to its weak result on sleep stage classification, where only two bi-polar EEG channels are available. The proposed Latent Alignment technique on the other hand still resulted in a significant increase in classification accuracy, and could therefore be more well-suited in cases where only very few electrodes are available.

In spite of the general increase in classification accuracy obtained when using Latent Alignment, this study also highlighted a weakness of statistical domain adaptation applied on the latent space under conditions of severe class imbalance. Domain adaptation under simultaneous feature and class distribution shift has been explored recently in other fields \cite{tan2020class, moreno2012unifying}. These approaches should be considered for incorporation into EEG decoding.

\section{Conclusion}
\label{sec:conclusion}

The present study confirmed that deep learning models trained on EEG signals without artifact removal can learn to utilize class-discriminative artifacts to increase classification accuracy. This was mitigated by inspecting the spatial filters of trained models as well as the raw signals themselves in order to remove contaminated parts of the trial. It is recommended to carefully evaluate trained deep learning models to avoid spurious classification results when conducting performance benchmarks.

We introduced the Latent Alignment approach to unifying distribution shifts between subjects in EEG decoding. A reformulation of Latent Alignment as a deep set architecture applied to sets of EEG trials was developed. As such, Latent Alignment incorporates the additional information available in the set rather than decoding each trial individually, which offers a different perspective on the benefits of subject distribution alignment.

Different subject adaptation techniques were compared against a fully subject-independent baseline, revealing that our proposed method of Latent Alignment achieved the best classification accuracy across various tasks and deep learning model architectures.

An analysis of the impact of class imbalance in the context set used by statistical alignment techniques revealed that aligning at later stages of the model, while improving classification accuracy, makes models more susceptible to class imbalance. This is an open problem that should be addressed in future research.


\bibliographystyle{cas-model2-names}
\bibliography{references}

\end{document}